\documentclass[12pt]{article} 
\usepackage{amsmath}
\usepackage{graphicx}
\usepackage{enumerate}
\usepackage{natbib}
\usepackage{multirow}
\usepackage{longtable}
\usepackage{booktabs}
\usepackage{amssymb}
\usepackage{bm}
\usepackage{mathrsfs}
\usepackage{amsthm}
\usepackage{amsfonts}
\usepackage{subfigure}
\usepackage{floatrow}
\usepackage{epsfig,amssymb,latexsym,verbatim}
\usepackage{epstopdf}
\usepackage{graphics}
\usepackage[english]{babel}
\usepackage[figuresright]{rotating}
\usepackage[dvipsnames]{xcolor}
\usepackage{color}
\usepackage{hyperref}

\usepackage{algorithm}
\usepackage{algpseudocode}

\usepackage{xr}

\allowdisplaybreaks
\newtheorem{theorem}{Theorem}

\newtheorem{definition}{Definition}
\newtheorem{assumption}{Assumption}
\newtheorem{corollary}{Corollary}
\numberwithin{equation}{section}
\numberwithin{lemma}{section}
\numberwithin{theorem}{section}
\numberwithin{prop}{section}
\numberwithin{corollary}{section}

\def\o{{\scriptstyle{\mathcal{O}}}}

\def\defeq{\stackrel{\mathrm{def}}{=}}

\newcommand{\blind}{0}

\begin{document}

\def\spacingset#1{\renewcommand{\baselinestretch}%
{#1}\small\normalsize} \spacingset{1}

\date{}
\if0\blind
{
  \title{\bf Federated Learning of Quantile Inference under Local Differential Privacy}
  \author{Leheng Cai\footnotemark[1] \footnotemark[4]$\,$ ,$\,$  Qirui Hu\footnotemark[2] \footnotemark[3] \footnotemark[4] $\,$ and$\,$ Shuyuan Wu\footnotemark[2] \footnotemark[4]\hspace{.2cm}}
  \maketitle
  \renewcommand{\thefootnote}{\fnsymbol{footnote}}

  \footnotetext[1]{Department of Statistics and Data Science, Tsinghua University}
   \footnotetext[2]{School of Statistics and Data Science, Shanghai University of Finance and Economics}
           
		\footnotetext[3]{Corresponding author: huqirui@mail.shufe.edu.cn}
        \footnotetext[4]{Equal contribution with alphabetical order}
} 

\if1\blind
{
  \bigskip
  \bigskip
  \bigskip
  \begin{center}
    { \bf   }
\end{center}
  \medskip
} \fi

\bigskip

\maketitle
	
	\textbf{Abstract: }   
In this paper, we investigate federated learning for quantile inference under local differential privacy (LDP). We propose an estimator based on local stochastic gradient descent (SGD), whose local gradients are perturbed via a randomized mechanism with global parameters, making the procedure tolerant of communication and storage constraints without compromising statistical efficiency. Although the quantile loss and its corresponding gradient do not satisfy standard smoothness conditions typically assumed in existing literature, we establish asymptotic normality for our estimator as well as a functional central limit theorem. The proposed method accommodates data heterogeneity and allows each server to operate with an individual privacy budget. Furthermore, we construct confidence intervals for the target value through a self‐normalization approach, thereby circumventing the need to estimate additional nuisance parameters. Extensive numerical experiments and real data application validate the theoretical guarantees of the proposed methodology.


	\maketitle
	
\newpage
\spacingset{1.9} 
\section{Introduction}
Quantile estimation and inference are critical tools in various scientific and applied domains. In healthcare, quantile methods facilitate more informed decisions regarding the optimal distribution of scarce medical resources, thus promoting equitable and effective patient care \citep{yadlowsky2025evaluating}. Similarly, quantile techniques have proven highly valuable in policy evaluation, as they reveal heterogeneous effects across different subgroups, nuances typically obscured by traditional average-based analyses \citep{kallus2024localized, chernozhukov2011inference, chernozhukov2005iv}. In reliability engineering, quantile-based approaches have significantly improved the assessment of system robustness, particularly under rare or extreme conditions, demonstrating their broad applicability and precision \citep{he2023smoothed,hu2022stochastic}. Moreover, finance widely employs quantile-based metrics such as value-at-risk, essential for managing financial risks in the face of regulatory pressures and market uncertainties \citep{barbaglia2023forecasting, chen2008nonparametric, wang2012estimation}. In general, quantile methods excel at capturing the characteristics of skewed or extreme-valued data, delivering richer insights into complex distributions prevalent in practical scenarios \citep{angrist2006quantile,chen2023recursive}.

Traditional quantile estimation methods have been extensively studied. However, with the rapid increase in massive datasets \citep{jordan2019communication,hector2021distributed,fan2023communication},  traditional approaches that rely on analyzing all data on a single machine may no longer be computationally feasible. This challenge has motivated the emergence of federated learning methods \citep{mcmahan2017communication,liu2020accelerating,tian2023unsupervised}. Federated learning enables multiple distributed clients to collaboratively train a global model without exchanging raw data, effectively addressing computational efficiency and privacy concerns \citep{konevcny2016federated}. 
In standard federated learning, a central server coordinates iterative model updates among clients, and under suitable conditions, this process guarantees convergence \citep{li2020federated,chen2022first}. To further enhance communication efficiency, local stochastic gradient descent (SGD) has been proposed, allowing clients to perform multiple local updates before synchronization. Under i.i.d. scenarios, the theoretical optimality of local SGD has been established \citep{stich2018local}. However, data heterogeneity, which frequently occurs in federated learning, significantly complicates local SGD. A series of studies have investigated this issue by analyzing convergence in worst-case heterogeneous scenarios \citep{hu2024does}, proposing regularization techniques to ensure local models remain close to the global model \citep{li2020federated}, and introducing momentum-based algorithms to stabilize training under non-i.i.d. conditions \citep{li2025federated}. Moreover, inference methods have also been developed and analyzed \citep{li2022statistical}.

\par Federated learning aggregates individual information to enable efficient collaborative model training. These personal data power indispensable services, from facial recognition unlocking our phones to recommendation systems curating news feeds, but they also carry latent risks. Leaked genetic markers can jeopardize insurance rates, and cleverly crafted prompts can coax large language models into regurgitating fragments of their private training corpora \cite{nasr2023scalable}. Differential privacy (DP) offers a principled defense: by bounding the statistical influence of any single participant, DP ensures that outputs remain virtually unchanged whether or not an individual opts in \cite{dwork2006our}. This safeguard, however, dissolves if the data custodian is breached, coerced, or simply misconfigured access controls—scenarios illustrated by repeated healthcare leaks and high-profile cloud missteps. Local differential privacy (LDP) fortifies the pipeline by introducing randomness at the point of collection: users perturb their data locally, send only noisy summaries, and retain the key to their raw information \cite{duchi2013local}. Even a fully compromised server therefore receives nothing decipherable. Industry adoption is accelerating: Google’s RAPPOR measures Chrome settings, iOS uses LDP to count emoji preferences, and Windows telemetry applies similar techniques to malware prevalence \cite{erlingsson2014rappor,ding2017collecting}. Collectively, these systems prove that granular user analytics and uncompromising privacy need not be mutually exclusive; instead, LDP sets a practical, legally robust baseline for responsible data-driven innovation.

\par DP federated learning has attracted considerable attention recently (e.g., \citep{liu2023novel, agarwal2018cpsgd, shi2022distributionally, ma2022differentially}). The additional communication layer between local clients and the global server gives rise to distinct privacy requirements. As delineated in \citep{lowy2023private}, one can categorize DP at the individual record level, inter‐silo record level, shuffled‐model, and user‐level, in order of increasing trust assumptions. In particular, LDP posits that each individual does not trust any other party, including their own silo, and therefore must randomize her report before release. Extensive work has focused on this setting (e.g., \citep{zhao2020local, shen2023pldp, jiang2022signds}).

\par Whereas prior studies of LDP in federated learning (e.g., \citep{zhao2020local, shen2023pldp, jiang2022signds}) primarily address estimation, statistical inference, such as constructing confidence intervals and conducting hypothesis tests, poses additional challenges. Beyond deriving the limiting distribution, inference requires a consistent estimator of the asymptotic variance. For SGD‐based methods, this typically involves the Hessian matrix, which exists only for smooth loss functions \citep{chen2020statistical}. Moreover, because only privatized gradients are observed, one may need extra privacy budget or data‐splitting to estimate variance reliably. Finally, existing single‐machine LDP quantile algorithms, such as \cite{huang2021instance} or \cite{liu2023online} cannot derive the inference result or do not readily extend to federated settings due to client‐heterogeneity in local loss functions.

\par In this paper, we propose a novel federated learning algorithm for quantile inference under LDP. Our method accommodates client‐level heterogeneity in quantile targets, privacy budgets, and data distributions, thereby enhancing the applicability of quantile inference in realistic federated environments.  A key innovation is our theoretical analysis of the local SGD quantile estimator. We first design an LDP mechanism that effectively reduces the federated quantile estimation problem to an equivalent non‐private setting. Exploiting this reduction, we establish the estimator’s asymptotic normality and derive a functional central limit theorem without average‐smoothness condition on the loss function. To the best of our knowledge, this constitutes the first weak‐convergence result for local SGD when the loss does not satisfy the usual average‐smoothness condition \citep{li2022statistical, xie2024asymptotic, zhu2024high}. Building on these non‐private asymptotic results, we develop an LDP‐compliant inference procedure for federated quantile estimation. By employing a self‐normalization technique, we avoid direct estimation of the asymptotic variance, instead constructing confidence intervals that automatically eliminate the unknown variance term. To the best of our knowledge, we provide the first inference framework for federated quantile estimation, even without privacy constraints.

\par The remainder of the paper is organized as follows. Section~\ref{Sec:method} reviews background and notation. Section~\ref{Sec:asym} presents the asymptotic analysis of the proposed estimator. Section~\ref{Sec:simu} reports extensive numerical experiments and real data application. All technical proofs and additional simulation results are deferred to the Appendix.

\section{Methodologies}\label{Sec:method}

First, we recall the definitions of central and local differential privacy. We then describe our problem setting and algorithmic details.

\begin{definition}[Central Differential Privacy, CDP \citep{dwork2006our}]
	A randomized algorithm $\mathcal{A}$ operating on a dataset $S$ is \emph{$(\epsilon,\delta)$-differentially private} if, for any pair of datasets $S$ and $S'$ differing in a single record and for any measurable set $E$,  
	\[
	\Pr\bigl[\mathcal{A}(S)\in E\bigr]\le e^{\epsilon}\,\Pr\bigl[\mathcal{A}(S')\in E\bigr]+\delta.
	\]
	When $\delta=0$, $\mathcal{A}$ is called \emph{$\epsilon$-DP}.
\end{definition}

\begin{definition}[Local Differential Privacy, LDP \citep{8948625}]
	A family of randomized mappings $R:X\to Y$ is \emph{$(\epsilon,\delta)$-locally differentially private} if, for every pair of inputs $x,x'\in X$ and every measurable subset $E\subseteq Y$,
	\[
	\Pr\bigl[R(x)\in E\bigr]\le e^{\epsilon}\,\Pr\bigl[R(x')\in E\bigr]+\delta.
	\]
\end{definition}

Under CDP, a trusted curator collects the raw data and adds noise before release; this model simplifies algorithm design and typically incurs only an $\mathcal{O}(1/n)$ loss in accuracy \citep{cai2021cost},  where $n$ denotes the sample size. In contrast, LDP dispenses with any trust assumption: each user $i$ holds a private value $X_i$, applies a predetermined randomized mechanism $R_i$ satisfying $(\epsilon,\delta)$-DP, and submits only the perturbed output. We adopt the non‐adaptive LDP model, in which all randomizers $\{R_i\}$ are fixed in advance \citep[Definitions~2.3 and~2.6]{cheu2019distributed}. Consequently, inference must proceed solely from locally privatized data.

In the CDP setting, the privatized estimator $\widehat\theta_{\mathrm{CDP}}$ typically satisfies  
$
\widehat\theta_{\mathrm{CDP}} - \theta = \mathcal{O}_p\bigl(n^{-1}\bigr),
$  
thus after $\sqrt{n}$‐scaling, it shares the same asymptotic distribution as the non‐private estimator, and one can recover its asymptotic variance with modest additional privacy cost. Under LDP, however, the error rate degrades to  
$
\widehat\theta_{\mathrm{LDP}} - \theta = \mathcal{O}_p\bigl(n^{-1/2}\bigr),
$  
which both alters the limiting law and inflates the asymptotic variance. Moreover, because only privatized data are available, consistently estimating this variance from data collected solely for point estimation is generally infeasible.

We consider a federated learning framework involving  $K$  clients, each independently holding a local dataset
i.i.d. drawn from an unknown distribution $\mathcal{P}_k$ with cumulative distribution function (CDF) $F_k$ and density function $f_k$ \citep{li2022statistical}. The goal is to collaboratively estimate the global quantile via weighted loss, i.e.,  the objective is to solve the following optimization problem:
\begin{align}\label{EQ:fedloss}
	\mathop{\arg\min}_{Q\in {\Theta} } \mathcal{L}(Q) \defeq  \mathop{\arg\min}_{Q\in {\Theta}} \sum_{k=1}^K p_k \mathcal{L}_{\tau_k}(Q) \defeq  \mathop{\arg\min}_{Q\in {\Theta}} \sum_{k=1}^K p_k \mathbb{E}_{x_k \sim \mathcal{P}_k}\{\ell_{\tau_k}(x_k,Q)\},
\end{align}
where $p_k$ denotes the weight assigned to client $k$, $\tau_k\in (0,1)$ is local quantile level, {\color{black}$x_k$ is the sample generated from $\mathcal{P}_k$}, and $\ell_{\tau_k}(x,Q)$ represents the check loss function defined as:
\begin{align}\label{EQ:localloss}
	\ell_{\tau_k}(x,Q) = (x - Q)(\tau_k - \mathbb{I}(x < Q)),
\end{align}
where $\mathbb{I}(\cdot)$ is the indicator function. Let $\tau:=\sum_{k = 1}^Kp_k\tau_k  \in (0,1)$. For the global minimizer $Q^{\star}$ of (\ref{EQ:fedloss}), it corresponds to the global quantile at level $\tau$ of a weighted CDF, i.e.,  $\sum_{k = 1}^Kp_kF_k(Q^{\star}) = \tau$. In the following, we denote $F_k(Q^{\star}) = Q_k$, and considers the parameter space ${\Theta}$ is bounded; see \cite{gu2023distributed}.

As noted in introduction, to improve the communication efficiency, we consider a local SGD based estimator, for communication iteration sets $\mathcal{I} = \{t_0,t_1,\dots,t_T\}$, the global server will receive the local iterations and broadcast the update to $K$ clients, otherwise, the iterations are only conducted in each local clients, i.e., for $k = 1,\dots K$, 
\begin{align*}
	q_{t+1}^k = \begin{cases} 
		q_{t}^k - \eta_t\left\{\mathbb{I}(x_{t}^k <q_t^k) - \tau_k\right\}, ~~ &t\notin \mathcal{I},\\
		\sum_{k = 1}^Kp_k\left[q_{t}^k - \eta_t\left\{\mathbb{I}(x_{t}^k <q_t^k) - \tau_k\right\}\right], ~~ &t\in \mathcal{I}.
	\end{cases}
\end{align*}
{\color{black} Here $\eta_t$ is the pre-determined learning rate, and $x_t^k$ represents an independent realization of $\mathcal{P}_k$,}
The final estimator is Polyak-Ruppert type, which averages the historical iterations,
\begin{align*}
	\widetilde{Q}_T = \frac{1}{t_T}\sum_{m = 1}^T \sum_{k = 1}^K p_k q_{t_m}^k.
\end{align*}

\par The communication and statistical efficiency are determined by the interval length $E_m := t_{m} - t_{m-1}$ for $m \in \mathbb{N}^{+}$.  If $E_m = 1$, the local clients must communicate with the global server at every iteration. In this scenario, the approach reduces to parallel SGD, which, as noted by \cite{li2022statistical}, may achieve the Cram\'{e}r-Rao lower bound and thus serve as an efficient estimator for certain smooth loss functions. Conversely, if $E_m = n$, implying only one communication at the last iteration, the estimator degenerates to a divide-and-conquer estimator. Such an estimator is consistent only when $\tau_k \equiv  \tau$ for all $k = 1,\dots, K$ and some common $\tau \in (0,1)$. In this case, minimizing the loss function (\ref{EQ:fedloss}) becomes a distributed learning problem. However, as pointed out by \cite{gu2023distributed}, the divide-and-conquer estimator may still be statistically inefficient for certain weight choices. Therefore, a careful balance must be struck between communication and statistical efficiency. For a general positive interval $E_m > 0$, the local SGD method allows us to find an appropriate choice of $E_m$ that can ensure an optimal trade-off between these efficiencies.

\par On the other hand, the data collected from each client may be subject to privacy protection policies, particularly in surveys involving sensitive information such as income or health status. For the local quantile loss function~\eqref{EQ:localloss}, we observe that the structure of its gradient resembles a binary response. This motivates us to incorporate an LDP mechanism based on randomized response and permutation, following the framework of \cite{liu2023online}, with a truthful response rate $r_k \in (0,1]$. Specifically, the mechanism allows each local client to either return a true gradient with probability $r_k$ or a synthetic Bernoulli random variable with probability $1 - r_k$. This iterative mechanism ensures $\epsilon_k$-LDP, where the privacy parameter is given by $\epsilon_k = \log(1 + r_k) - \log(1 - r_k)$, as established in \cite{liu2023online}. 

\par
\begin{figure}[h!]
	\centering
	\subfigure[Data heterogeneity]{%
		\includegraphics[width=0.4\linewidth]{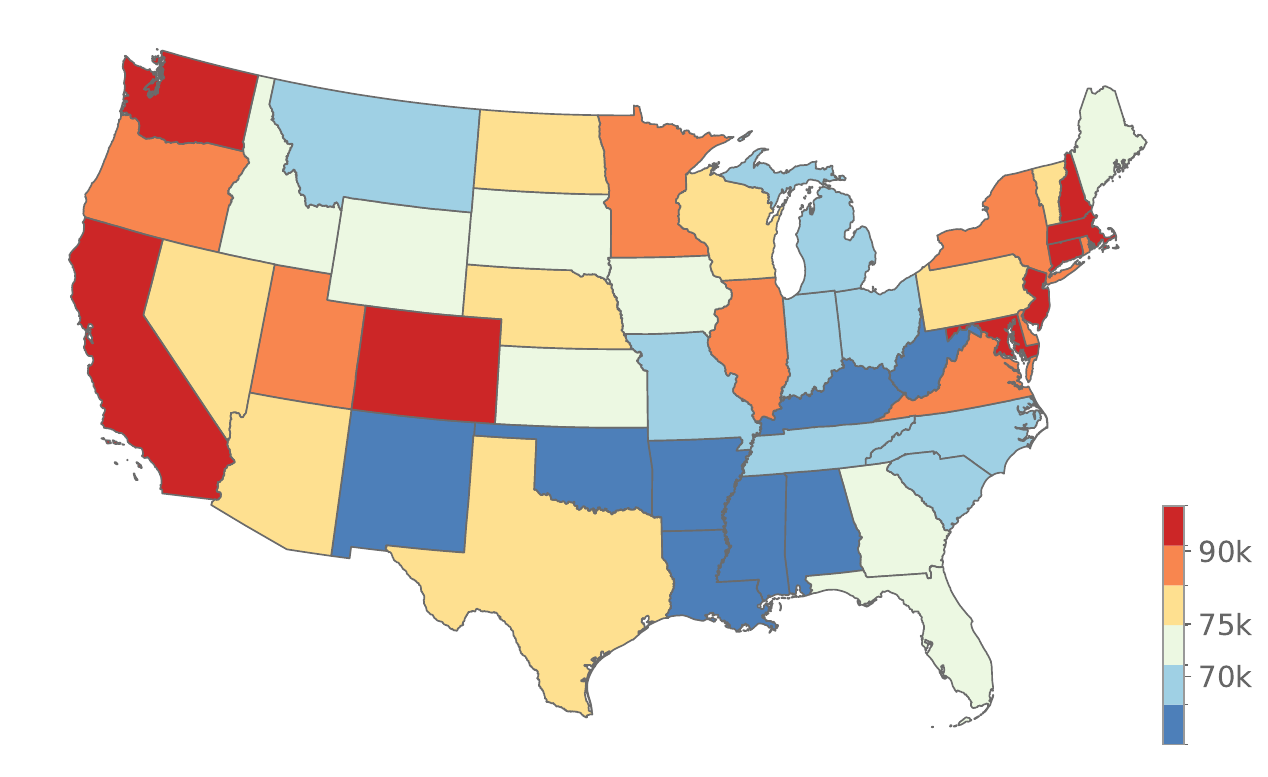}}
	\hfill
	\subfigure[Privacy heterogeneity]{%
		\includegraphics[width=0.58\linewidth]{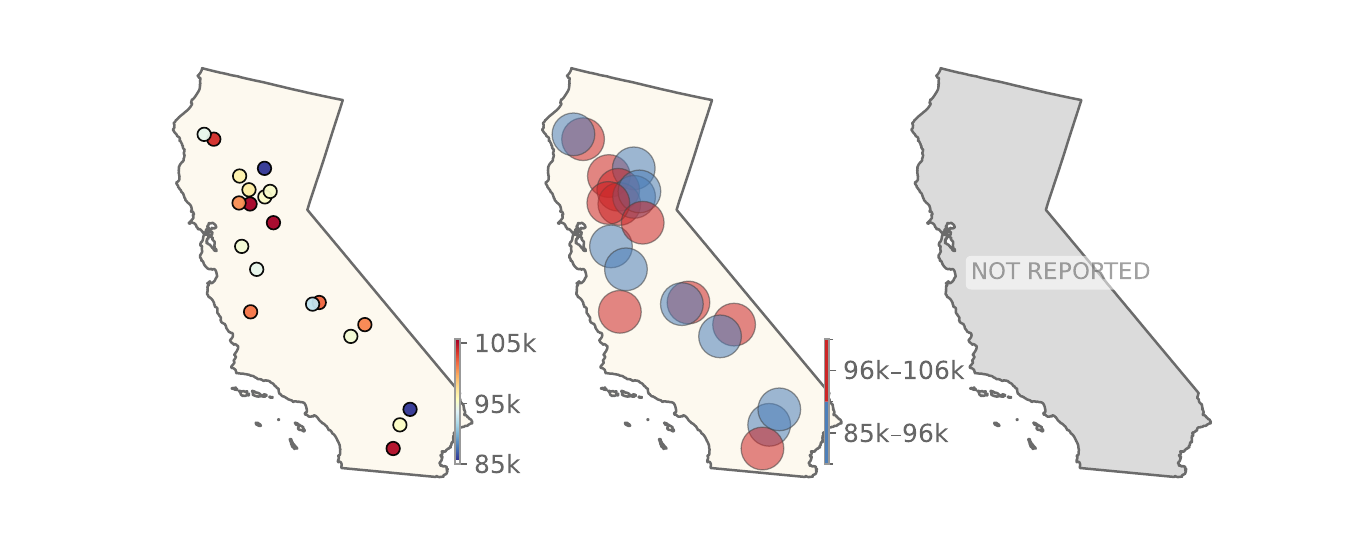}}\caption{Illustration of client heterogeneity. Income data source: U.S. Census Bureau (\url{https://data.census.gov/table/ACSST5Y2023.S1901?g=010XX00US0400000}). Panel (i) shows median annual income by state. Panel (ii) shows three income disclosure schemes under different privacy budgets: (a) each individual release true income; (b) each individual release an income interval; and (c) withholding release.}
	\label{fig:example}
\end{figure}

 It is worth noting that the method of \cite{liu2023online} was originally developed in a single-machine setting. Extending it directly to federated learning raises new challenges, since federated systems inherently involve the issue of heterogeneity.  
We illustrate with a simple example. Consider collaboratively estimating the national median annual income using state-level data from the United States, where each state is treated as a client. First, income distributions typically vary across states (see Figure~\ref{fig:example}(i)). Second, privacy preferences can differ across states due to cultural norms and development levels \citep{milberg2000information,bellman2004international}. Figure~\ref{fig:example}(ii) shows how the released information can vary under different privacy budgets. Due to such heterogeneity, a naive combination of local LDP estimators from \cite{liu2023online} may result in severely biased results.
To address this problem, we propose a novel federated quantile estimation algorithm under LDP, equipped with a carefully designed local SGD updating rule. This method accommodates heterogeneous data distributions, quantile targets, and privacy budgets across clients while maintaining a common global target. The complete procedure is summarized in Algorithm~\ref{alg:main}, and we denote the resulting estimator as $\widehat{Q}_T$.

\begin{algorithm}[H]
	\raggedright
	\caption{Federated quantile estimation with local SGD under LDP}
	\label{alg:main}
	\textbf{Input:} step sizes $\{\eta_m\}_{m=1}^T$, target quantile $\tau \in (0,1)$, truthful response rates $\{r_k\}_{k=1}^K$, communication set $\mathcal{I} = \{t_0,t_1,\dots,t_T\}$. 
	\\
	\textbf{Initialization:} set $q_0^k = q_0 \sim \mathcal{N}(0,1)$  for all $1 \leq k \leq K$, let  $\widehat{Q}_0\leftarrow0$.
	\begin{algorithmic}
		\For{$m = 1$ to $T$}
		\For{$k=1$ to $K$ (distributedly)}
		\For{$t = t_{m-1} + 1$ to $t_m$}  \Comment{Local updates}
		\State $u_t^k \sim Bernoulli(r), \quad v_t^k \sim Bernoulli(0.5)$ 
		\State $s_t^k = \mathbb{I}(x_t^k>q_{t-1}^k)\mathbb{I}(u_t^k=1) + v_t^k\mathbb{I}(u_t^k=0)$
		\State {\color{black} $q_{t}^k \leftarrow q_{t-1}^k + \dfrac{1 - r_k + 2\tau r_k}{2r_k}\eta_{m-1}\mathbb{I}(s_t^k = 1) - \dfrac{1 + r_k - 2\tau r_k}{2r_k}\eta_{m-1}\mathbb{I}(s_t^k=0)$ }
		\EndFor
		\EndFor
		\State $\bar{q}_{t_m} \leftarrow \sum_{k=1}^K p_k q_{t_m}^k$; $q_{t_{m}}^k \leftarrow \bar{q}_{t_{m}}$ for all $1 \leq k\leq K$. \Comment{Aggregation and synchronization.}
		\State $\widehat{Q}_m \leftarrow ((m-1) \widehat{Q}_{m-1} + \bar{q}_{t_m})/m$.
		\EndFor
	\end{algorithmic}
	\textbf{Return:} $\widehat{Q}_T$.
\end{algorithm}
In our proposed Algorithm \ref{alg:main}, each iteration integrates global information (the global quantile $\tau$) with client-specific privacy budget ($r_k$), thereby correcting bias arising from the aggregation of heterogeneous local LDP mechanisms and loss functions. The following theorem shows that this procedure effectively reduces the LDP inference problem to its non-private analogue.

\begin{theorem}\label{THM:LDPvaild}
	Denote $\widetilde{\tau}_k = r_k \tau + (1 - r_k)/2$. For a privacy budget $\epsilon_k = \log(1 + r_k) - \log(1 - r_k)$, there exists a dataset 
	consisting of i.i.d.\ samples drawn from some distribution $\widetilde{\mathcal{P}}_k, 1\leq k \leq K$, such that solving the federated loss~\eqref{EQ:fedloss} with $\epsilon_k$-LDP using data 
	drawn from $\mathcal{P}_k$ is equivalent to solving the following non-private problem:
	\begin{align}\label{EQ:DPfedloss}
		\mathop{\arg\min}_{Q} \mathcal{L}(Q)
		= \mathop{\arg\min}_{Q} \widetilde{\mathcal{L}}(Q) 
		:= {\color{black} \mathop{\arg\min}_{Q} \sum_{k=1}^K p_k \mathbb{E}_{x_k \sim \widetilde{\mathcal{P}}_k}\left\{r_k^{-1}\ell_{\widetilde{\tau}_k}(x_k, Q)\right\}.}
	\end{align}
\end{theorem}

\par Therefore, by Theorem~\ref{THM:LDPvaild}, the LDP federated quantile estimation problem can be reformulated as a non-DP federated quantile estimation task under modified distributions and shifted quantile levels. The main challenge then becomes analyzing the statistical properties of the resulting non-DP estimator, particularly in the presence of the non-smooth quantile loss function.

\section{Asymptotic analysis}\label{Sec:asym}
\par In this section, we focus on the asymptotic analysis of the proposed LDP estimator and the practical construction of confidence intervals. Before presenting the main results, we first introduce several necessary assumptions.

\begin{assumption} \label{ass:bound}
      For some   constant $C_{f }>0$,  $ f_k (\cdot) $, $1\leq k\leq K$, is uniformly bounded by $C_{f }$.
\end{assumption}
\begin{assumption}\label{ass:sstepsize}
     Define the effective step $\gamma_m=\eta_m E_m$, which is non-increasing in $m$ and satisfies that $\sum_{m=1}^{\infty}\gamma_m^2<\infty$, $\sum_{m=1}^{\infty}\gamma_m=\infty$, and $ {(\gamma_m-\gamma_{m+1})}/{\gamma_m}=\o(\gamma_m).$ 
\end{assumption}
\begin{assumption}\label{ass:Em}
     The sequence $\{E_m\}_{m\geq 1}$ satisfies that  
\begin{itemize}
    \item[(a)] $\{E_m\}_{m\geq 1}$ is either uniformly bounded or non-decreasing.
    \item[(b)] There exist some $\delta>0$ and  $\nu\geq 1$ such that \begin{align*}
        \limsup_{T\to\infty}\frac{1}{T^2}\left(\sum_{m=0}^{T-1}E_m^{1+\delta}\right)\left(\sum_{m=0}^{T-1}E_m^{-1-\delta}\right)<\infty,  \lim_{T\to\infty}\frac{1}{T^2}\left(\sum_{m=0}^{T-1}E_m \right)\left(\sum_{m=0}^{T-1}E_m^{-1 }\right)=\nu.
    \end{align*}
    \item[(c)] Denote $t_T=\sum_{m=0}^{T-1}E_m$, satisfying  \begin{align*}
        \lim_{T\to\infty}\frac{
        \sqrt{t_T}
        }{T} \sum_{m=0}^{T}\gamma_m = 0,\quad \lim_{T\to\infty}\frac{
        \sqrt{t_T}
        }{T}\frac{1}{\sqrt{\gamma_T}}=0
    \end{align*}
\end{itemize}
\end{assumption}
Assumption 1 is a mild and regular condition concerning the uniform boundedness of density functions. Assumptions 2 and 3 require that the effective step sizes decay slowly and the communication intervals increase slowly; see also \cite{li2022statistical}. 

\begin{theorem}\label{THM:CLT}
\par Under Assumptions 1-3, as $T\to \infty$, the proposed LDP federated estimator enjoys
 \begin{align*}
    \sqrt{t_T}(\widehat{Q}_T - Q^{\star}) \xrightarrow{d} N\left(0, {\color{black}\nu\frac{\sum_{k = 1}^Kp_k^2\left\{r_k^{-2}-(2Q_k-1)^2\right\} }{4\left(\sum_{k=1}^K p_kf_k(Q^{\star})\right)^2}}\right).
\end{align*}

\end{theorem}


\par Theorem \ref{THM:CLT} establishes the asymptotic normality of the estimator $\widehat Q_T$, which theoretically  allows for the theoretical construction of a confidence interval for $Q^{\star}$. However, the construction involves unknown quantities, such as the individual quantiles $Q_k$ and the density values $f_k(Q^{\star})$. Even in cases where $Q_k = \tau_k$ is known,  the estimation of $f_k(Q^{\star})$ remains challenging. In particular, it is difficult to recover these density values using only the perturbed gradients available from Algorithm~\ref{alg:main}. Moreover, in SGD-based methods, consistent variance estimation typically relies on the Hessian matrix, which is well-defined only for smooth loss functions, as previously discussed. Therefore, although Theorem~\ref{THM:CLT} provides a theoretically valid basis for confidence interval construction, it is not practically implementable due to these limitations.

\par Inspired by the quantile inference framework for single clients in \cite{liu2023online}, it is necessary to strengthen the pointwise result of Theorem~\ref{THM:CLT} to a functional version.

\begin{theorem}\label{THM:FCLT}
Under Assumptions 1–3, as $T \to \infty$, we have
\[
\mathcal{Q}_T({\color{black}s}) := \frac{\sqrt{t_T}}{T} \sum_{m=1}^{h({\color{black}s}, T)} \left(\bar{q}_{t_m} - Q^{\star} \right) \xrightarrow{d}  {\color{black}\frac{\sqrt{\nu\sum_{k = 1}^Kp_k^2\left\{r_k^{-2}-(2Q_k-1)^2\right\}}}{2\sum_{k=1}^K p_kf_k(Q^{\star})}B(s)},
\]

where $t_T = \sum_{m=0}^{T-1} E_m$, $\bar{q}_{t_m} = \sum_{k=1}^K p_k r_k q_{t_m}^k$, $B(\cdot)$ is a standard Brownian motion on $[0,1]$, and
\[
h({\color{black}s}, T) = \max \left\{ n \in \mathbb{Z}_{>0} \,\middle|\, {\color{black}s} \sum_{m=1}^T \frac{1}{E_m} \geq \sum_{m=1}^n \frac{1}{E_m} \right\}, \quad \text{for } {\color{black}s} \in (0,1].
\]
\end{theorem}

Theorem \ref{THM:FCLT} establishes a functional central limit theorem (FCLT) for $ Q_T(s)$ over $s \in (0,1]$, showing that it converges weakly in the $\ell^\infty[0,1]$ (the space of bounded real-valued functions) to a Brownian motion, {\color{black}which is our another  major theoretical contribution. Note that the sample quantile loss doesn't satisfy the common $L$-average smooth conditions for weakly convergence result, such in \citep{li2022statistical, xie2024asymptotic, zhu2024high}, leading to extra challenge in deriving the almost sure and $\mathcal L^2$ convergence rates of $\bar q_{t_m}$, which are  essential for handling the asymptotically negligible terms.} Theorem \ref{THM:CLT} arises as a special case of Theorem \ref{THM:FCLT} when $s = 1$. Building on Theorem \ref{THM:FCLT}, we proceed to construct a self-normalized test statistic and derive its asymptotic pivotal distribution via the continuous mapping theorem.

\par 
Define $r_0 = 0$ and, for $m \geq 1$,
$
r_m = \left( \sum_{i=1}^m 1/E_i \right) \left( \sum_{i=1}^T 1/E_i \right)^{-1},
$
which ensures that
\[
\mathcal{Q}_T(r_m) = \frac{\sqrt{t_T}}{T} \sum_{i=1}^m \left( \bar{q}_{t_i} - Q^{\star} \right), \,\,\, \text{and in particular, }\,\,\, \mathcal{Q}_T(1) = \frac{\sqrt{t_T}}{T} \sum_{i=1}^T \left( \bar{q}_{t_i} - Q^{\star} \right).
\]
 Following the arguments in \citep{shao2015self}, once a functional central limit theorem such as Theorem~\ref{THM:FCLT} is established, one can construct a self-normalized statistic that asymptotically enjoys a   pivotal   distribution. Specifically, define
\begin{align}\label{EQ:SNer}
\mathcal{V}_T = \sum_{m=1}^T \left( r_m - r_{m-1} \right) \left( \mathcal{Q}_T(r_m) - \frac{m}{T} \mathcal{Q}_T(1) \right)^2.
\end{align}

\begin{corollary}\label{COR:SN}
Suppose Assumptions 1-3 hold and $g(r_m) \asymp m/T$ for some continuous function $g$ on $[0,1]$. Then, as $T \to \infty$,
\[
\frac{\mathcal{Q}_T(1)}{\sqrt{\mathcal{V}_T}} \xrightarrow{d} \frac{B(1)}{\sqrt{\int_0^1 \left( B(r) - g(r) B(1) \right)^2 dr}}.
\]
\end{corollary}

Corollary \ref{COR:SN} presents the asymptotic distribution of the self-normalized statistic $Q_T(1)/\mathcal V_T$, which is distribution-free. As a result, there is no need to allocate additional DP budget to estimate nuisance parameters when constructing confidence intervals.

\par The selection of the self-normalizer is not unique, and an appropriate norm of the Gaussian process $B(r) - g(r) B(1)$ can yield similar results to those in Corollary~\ref{COR:SN}. For example, using the supremum norm  and the $\mathcal L_1$ norm, one can define alternative self-normalizers as follows:
\begin{align*}
    \mathcal{V}_T^{\prime} = \sup_{1 \leq m \leq T} \left| \mathcal{Q}_T(r_m) - \frac{m}{T} \mathcal{Q}_T(1) \right|, \quad
    \mathcal{V}_T^{\prime\prime} = \sum_{m=1}^T \left( r_m - r_{m-1} \right) \left| \mathcal{Q}_T(r_m) - \frac{m}{T} \mathcal{Q}_T(1) \right|,
\end{align*}
which are related to the processes $\sup_{0 \leq r \leq 1} |B(r) - g(r) B(1)|$ and $\int_{0}^1 |B(r) - g(r) B(1)| \, dr$, respectively.  However, the self-normalizer defined in equation~\eqref{EQ:SNer} enjoys greater computational efficiency, as the $\mathcal L_2$ norm can be computed in an online manner, as described in Algorithm~\ref{alg:inference}. Let $\widehat{\mathcal{V}}_T$ denote the estimator of the self-normalizer in~\eqref{EQ:SNer}, and let $v_{\alpha/2, g}$ be the $(1 - \alpha/2)$ quantile of the random variable 
$B(1) /\left(\int_0^1\left(B(r)-g(r) B_1(1)\right)^2 d r\right)^{1 / 2}$.
The following corollary ensures the asymptotic validity of the constructed LDP confidence interval.

The following Corollary \ref{COR:SNinterval} ensures the asymptotic validity of the constructed LDP confidence interval.

\begin{corollary}\label{COR:SNinterval}
Suppose the same conditions in Theorem \ref{THM:FCLT} hold, as $T\to \infty$, one has that 
\begin{align*}
    \mathbb{P}\left(\widehat{Q}_{T}-v_{\frac{\alpha}{2}, g} \sqrt{\widehat{\mathcal{V}}_{T}} \leq Q^{\star}\leq \widehat{Q}_{T}+v_{\frac{\alpha}{2}, g} \sqrt{\widehat{\mathcal{V}}_{T}} \right) \rightarrow 1-\alpha
\end{align*}
\end{corollary}

\begin{algorithm}[H]
\caption{Online Inference }
\raggedright
\label{alg:inference}
\textbf{Input:} step sizes $\{\eta_m\}_{m=1}^T$, target quantile $\tau \in (0,1)$, truthful response rates $\{r_k\}_{k=1}^K$, communication set $\mathcal{I} = \{t_0,t_1,\dots,t_T\}$.
\\
\textbf{Initialization:} set $q_0^k \sim \mathcal{N}(0,1)$  for all $k$, let $\mathcal{V}_0^a \leftarrow 0$, $\mathcal{V}_0^b \leftarrow 0$, $\mathcal{V}^s_0 \leftarrow0, \mathcal{V}^p_0 \leftarrow 0$, and $Q_0\leftarrow0$.
\begin{algorithmic}
\For{$m = 1$ to $T$}
\State Obtain $\widehat{Q}_m$ from Algorithm \ref{alg:main}.
\State $\mathcal{V}_m^a \leftarrow \mathcal{V}_{m-1}^a + m^2 Q_m^2/E_m$, \Comment{$E_m = t_m - t_{m-1}$}
\State $\mathcal{V}_m^b \leftarrow \mathcal{V}_{m-1}^b + m^2 Q_m/E_m$,
\State $\mathcal{V}_m^s \leftarrow \mathcal{V}_{m-1}^s + 1/E_m$,
\State $\mathcal{V}_m^p \leftarrow \mathcal{V}_{m-1}^p + m^2/E_m$.
\State 
$ \widehat{\mathcal{V}}_m \leftarrow \frac{1}{m^2 \mathcal{V}^s_m} \left( \mathcal{V}^a_m - 2  \mathcal{V}^b_m Q_m + \mathcal{V}^p_m Q_m^2 \right).
$ \Comment{Online inference.}
\EndFor
\end{algorithmic}
\textbf{Return:} Confidence interval $\big[\widehat{Q}_T-v_{\frac{\alpha}{2}, g} \sqrt{\widehat{\mathcal{V}}_T}, \ \  \widehat{Q}_T+v_{\frac{\alpha}{2}, g} \sqrt{\widehat{\mathcal{V}}_T}\big]$.
\end{algorithm}

\section{Experiments}\label{Sec:simu}
\subsection{Simulation setup}
We first evaluate our proposed method through extensive simulation studies using synthetic data. In all experiments, we fixed $p_k = 1/K$ for $1 \leq k \leq K$, the number of clients is fixed at $K = 10$. The quantile levels examined range from $0.3$ to $0.8$, and the truthful response rates vary between $0.25$ and $0.9$. We focus on the following four scenarios of heterogeneity:
\begin{itemize}
\item \textbf{heterogeneous quantile levels:} We investigate two distinct scenarios: (1) Case  $\tau_{\operatorname{low}}$: lower quantile levels, where each client is assigned a unique quantile level $\tau_k$ ranging uniformly from $0.3$ to $0.5$; and (2) Case $\tau_{\operatorname{high}}$: higher quantile levels, where $\tau_k$ ranges uniformly from $0.5$ to $0.8$. 
\item \textbf{heterogeneous response rates.} Each client has a unique truthful response rate $r_k$, ranging uniformly from 0.25 to 0.9. 
\item \textbf{heterogeneous locations (Hete L).} Data for each client $k$ are independently generated from $\mathcal{N}(\mu_k, 1)$, where  $\mu_k \sim \mathcal{N}(0,1)$. 
\item \textbf{heterogeneous distribution families (Hete D).} Data are generated independently across ten clients, with three drawing from $\mathcal{N}(0,1)$, three from the uniform distribution $\mathcal{U}(-1,1)$, and four from a standard Cauchy distribution $\mathcal{C}(0,1)$.
\end{itemize}
{\color{black} We set the step size $\gamma_m$ as: $\gamma_m = 20 \bar{r} / (m^{0.51} + 100)$, with $\gamma_m = E_m \eta_m$ and $\bar{r} = K^{-1} \sum_{k=1}^K r_k$.} 
Following \cite{li2022statistical}, we implement a warm-up phase, setting the communication interval $E_m = 1$ for the first 5\% of iterations. After the warm-up period, we redefine the interval sequence $\{E_m\}$ based on a new sequence $\{E_m^\prime\}$, specifically:
$E_m = E_{m - 0.05\cdot T}^{\prime}$. We examine three different interval strategies for $E_m^\prime$: (1) C1: $E_m^\prime \equiv 1$  (equivalent to P-SGD),
(2) C5: $E_m^\prime \equiv 5$, and 
(3) Log: $E_m^\prime = \lceil \log_2(m + 1) \rceil$. The initial parameter estimates are set to $q_0^k=q_0 \sim \mathcal{N}(0,1)$ for all clients $k$. 
All experimental settings are replicated $R = 1,000$ times. The simulations are conducted on computational resources comprising 36 Intel Xeon Gold 6271 CPUs, with a total of 128GB RAM and 500GB storage.  

\subsection{Simulation results}
\begin{figure}[h!]
  \centering
    \includegraphics[width=0.9\linewidth]{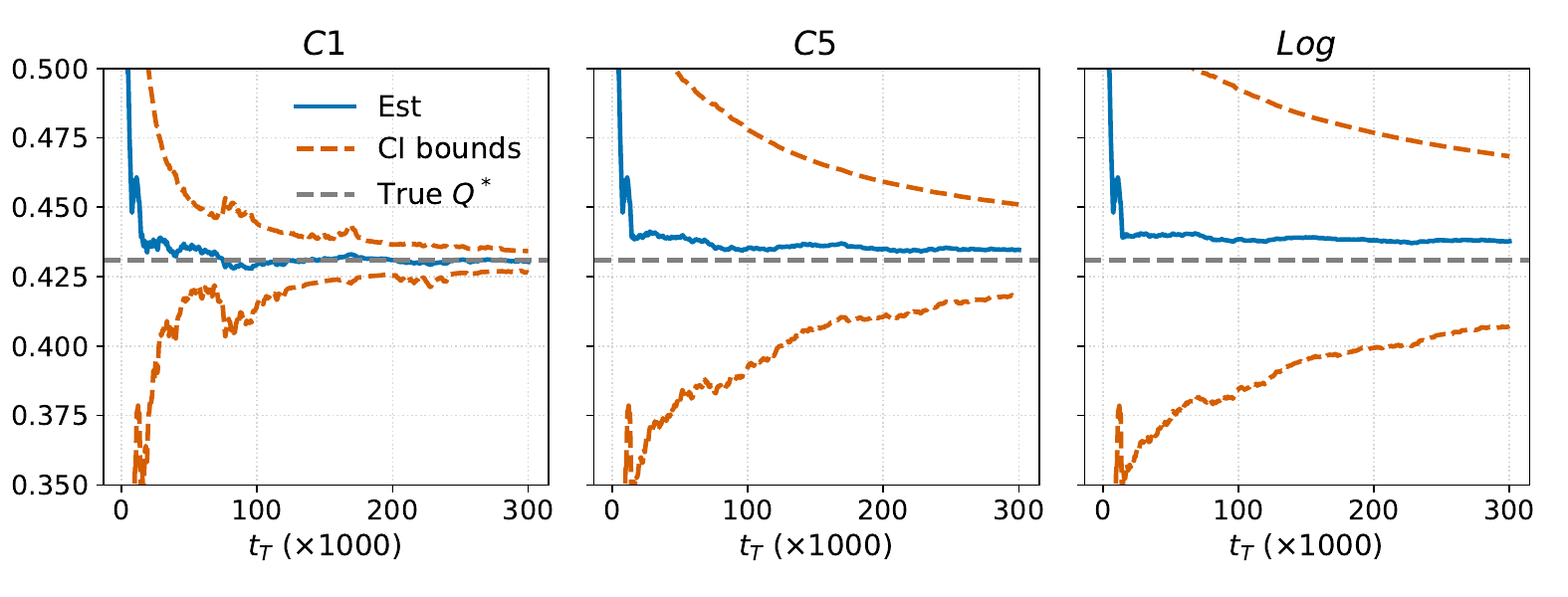}
  \hfill
    \includegraphics[width=0.9\linewidth]{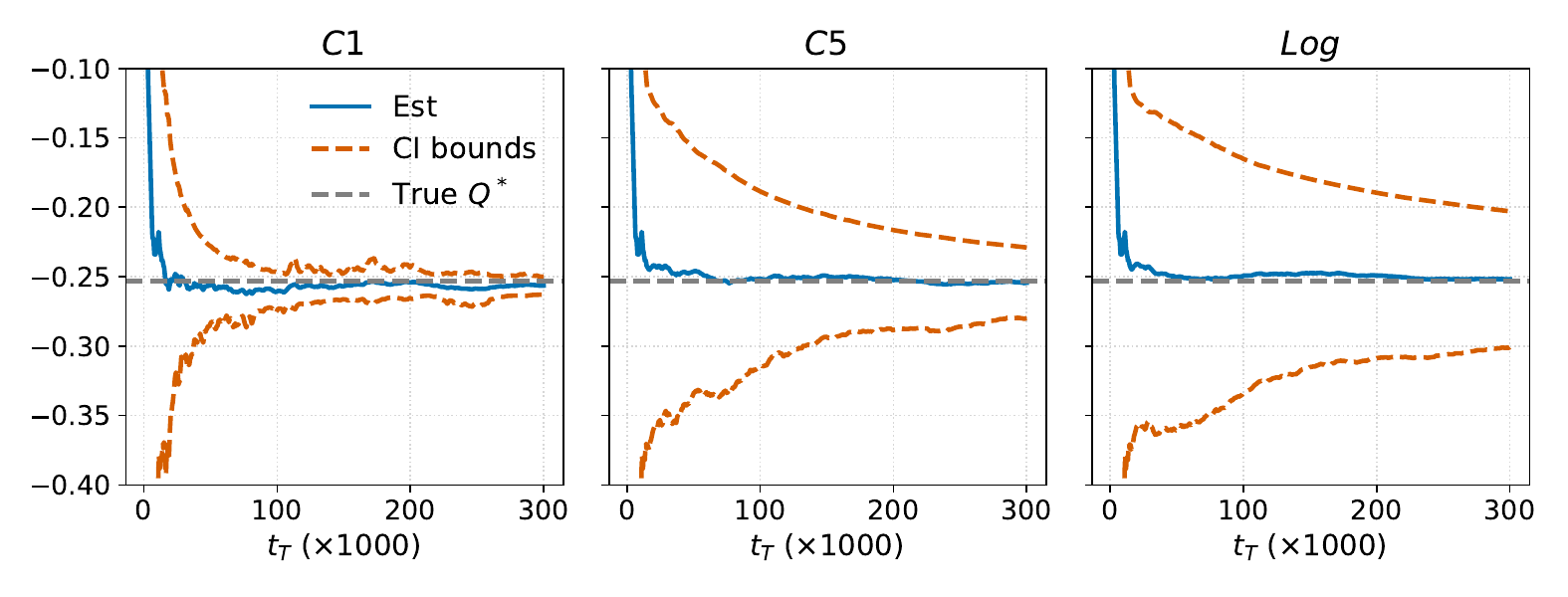}
  \caption{Sample trajectories of the iterative estimator and corresponding confidence intervals under heterogeneous distributions (Hete L, with $r_k=0.9$ and $\tau=0.5$, left panel) and heterogeneous quantile levels ($\tau_{\operatorname{low}}$, with heterogeneous response rates, right panel). The horizontal dotted line indicates the true quantile value $Q^*$.}
  \label{fig:trajec}
\end{figure}

We first illustrate the performance of our proposed method by presenting sample iteration trajectories for estimation and inference. Specifically, we randomly select one simulation run and plot the resulting estimates and corresponding confidence intervals against $t_T$ (Figure \ref{fig:trajec}). The results demonstrate that our approach accurately captures the true quantile value and provides reliable inference. Subsequently, we fix the total sample size $t_T$ at $10,000$ and $50,000$ and evaluate the finite sample performance under different settings. Let $\widehat{Q}^{(r)}_T$ denote the quantile estimator and $\operatorname{CI}^{(r)}$ represent the corresponding 95\% confidence interval obtained from Algorithm \ref{alg:inference} in the $r$-th simulation. We consider two metrics: the mean absolute error (MAE), defined as $R^{-1}\sum_{r=1}^{R} |\widehat{Q}^{(r)}_T - Q^*|$, and the empirical coverage probability (ECP), defined as $R^{-1}\sum_{r=1}^{R}\mathbb{I}(Q^*\in \operatorname{CI}^{(r)})$.  For comparison, {\color{black} we also consider two alternative methods:
(1) the DP-SGD method \citep{song2013stochastic}, which adds noise directly to the gradients instead of introducing DP through randomized response. To align with the original paper's setup, we focus on the case with $C=1$. In this regime, the gradient‐descent update in Algorithm \ref{alg:main} becomes
$$q_{t}^k \;\leftarrow\; q_{t-1}^k
+  \eta_{t-1} \Big\{ \tau_k\mathbb{I}(x_t^k>q_{t-1}^k) - (1 - \tau_k)\mathbb{I}(x_t^k<q_{t-1}^k) +Z_t^k \Big\},$$
where $Z_t^k$ is drawn from a Laplace distribution. A simple calculation shows that $Z_t^k$ has mean zero and scale parameter
$1/\log \left\{(1+r_k)/(1-r_k)\right\}$.  
(2) the divide-and-conquer (DC) method, which corresponds to the special case $E_m = n$. Here we use step size $\eta_t = 2 \bar{r} / (t^{0.51} + 100)$ \citep{goyal2017accurate}.  The numerical results for all of the methods are reported in Tables \ref{tab:fixed n vary tau} and \ref{tab:fixed n vary mu}.}

From Tables \ref{tab:fixed n vary tau} and \ref{tab:fixed n vary mu}, we observe that our method consistently achieves ECP close to or exceeding the nominal 95\% level across all scenarios. As either the total sample size $t_T$ or the truthful response rate increases, the MAE decreases, which aligns with our theoretical results. Comparing the three interval strategies, we find that the C1 strategy (P-SGD) yields the smallest MAE, as it has the highest communication frequency. {\color{black} Comparing with the two competing methods, we find that} the DC approach results in the largest errors. Notably, in certain heterogeneous cases, such as Hete L with $\tau = 0.8$, the DC estimator exhibits significant bias and an ECP far below the nominal 95\% level.  In contrast, our proposed estimators successfully achieve approximately 95\% empirical coverage in these cases. {\color{black} Moreover, while DP-SGD attains empirical coverage probabilities close to or even exceeding 95\% in most settings, its MAE remain uniformly larger than those of our method.}
To further illustrate the communication efficiency of our method, we also consider scenarios with a fixed number of communication rounds $T$.  The results are summarized in Tables \ref{tab:fixed T vary tau} and \ref{tab:fixed T vary loc}. We observe that our proposed method continues to provide valid inference. Additionally, under fixed communication rounds, the Log strategy generally achieves the best performance, yielding the smallest MAE.


 
\begin{table}[htbp]
\centering
\caption{Empirical coverage probabilities (mean absolute errors) under varying quantile levels and response rates, with different $t_T$ and fixed $K = 10$ clients and data generated from $\mathcal{N}(0,1)$. In Case $\tau_{\operatorname{low}}$, each client uses a unique quantile level $\tau_k$ ranging uniformly from $[0.3, 0.5]$; in Case $\tau_{\operatorname{high}}$, $\tau_k$ is ranging from $[0.5, 0.8]$. ``hetero'' indicates client-specific truthful response rates $r_k$ range from $[0.25, 0.9].$}
\resizebox{1\textwidth}{!}{
\begin{tabular}{ccccccc}
\toprule
Quantile ($\tau$) & Rate ($r$) & C1 & C5 & Log & DP-SGD (C1) & DC \\
\midrule
\multicolumn{7}{c}{$t_T = 10000$} \\
0.5 & 0.25 & 0.949(0.0133) & 0.967(0.0244) & 0.992(0.0360) & 0.949(0.0191)& 0.939(0.2503)\\
0.5 & hetero & 0.963(0.0071) & 0.989(0.0112) & 0.997(0.0161) & 0.955(0.0100)& 1.000(0.0497)\\
0.5 & 0.9 & 0.995(0.0023) & 1.000(0.0054) & 1.000(0.0082) & 0.980(0.0036)& 1.000(0.0158)\\
\hline
$\tau_{\operatorname{low}}$ & 0.25 & 0.947(0.0136) & 0.982(0.0253) & 0.990(0.0369) & 0.940(0.0200)& 0.969(0.2616)\\
$\tau_{\operatorname{low}}$ & hetero & 0.962(0.0072) & 0.993(0.0113) & 0.997(0.0162) & 0.949(0.0105)& 0.999(0.0530)\\
$\tau_{\operatorname{low}}$ & 0.9 & 0.999(0.0020) & 1.000(0.0055) & 1.000(0.0083) & 0.985(0.0036)& 1.000(0.0162)\\
\hline
$\tau_{\operatorname{high}}$ & 0.25 & 0.939(0.0145) & 0.987(0.0268) & 0.986(0.0399) & 0.952(0.0210)& 0.984(0.2771)\\
$\tau_{\operatorname{high}}$ & hetero & 0.968(0.0076) & 0.987(0.0126) & 0.999(0.0182) & 0.956(0.0111)& 1.000(0.0516)\\
$\tau_{\operatorname{high}}$ & 0.9 & 0.996(0.0023) & 1.000(0.0067) & 1.000(0.0102) & 0.980(0.0038)& 1.000(0.0172)\\
\hline
\multicolumn{7}{c}{$t_T = 50000$} \\
0.5 & 0.25 & 0.956(0.0056) & 0.982(0.0081) & 0.996(0.0122) & 0.949(0.0081)& 0.988(0.0571)\\
0.5 & hetero & 0.960(0.0032) & 0.979(0.0044) & 0.992(0.0064) & 0.950(0.0046)& 1.000(0.0115)\\
0.5 & 0.9 & 1.000(0.0018) & 0.988(0.0027) & 0.990(0.0036) & 0.983(0.0021)& 1.000(0.0038)\\
\hline
$\tau_{\operatorname{low}}$ & 0.25 & 0.957(0.0061) & 0.981(0.0083) & 0.994(0.0125) & 0.944(0.0091)& 0.993(0.0594)\\
$\tau_{\operatorname{low}}$ & hetero & 0.953(0.0036) & 0.981(0.0046) & 0.990(0.0066) & 0.934(0.0054)& 0.999(0.0121)\\
$\tau_{\operatorname{low}}$ & 0.9 & 1.000(0.0019) & 1.000(0.0026) & 0.989(0.0038) & 0.988(0.0024)& 1.000(0.0057)\\
\hline
$\tau_{\operatorname{high}}$ & 0.25 & 0.968(0.0059) & 0.986(0.0086) & 0.997(0.0133) & 0.946(0.0089)& 0.999(0.0620)\\
$\tau_{\operatorname{high}}$ & hetero & 0.953(0.0032) & 0.990(0.0045) & 0.998(0.0065) & 0.952(0.0047)& 0.993(0.0154)\\
$\tau_{\operatorname{high}}$ & 0.9 & 0.998(0.0010) & 0.999(0.0023) & 1.000(0.0034) & 0.977(0.0016)& 0.938(0.0132)\\
\bottomrule
\end{tabular}
}
\label{tab:fixed n vary tau}
\end{table}

\begin{table}[htbp]
\centering
\caption{Empirical coverage probabilities (mean absolute errors) under heterogeneous distributions for different $t_T$. The number of clients $K$ is fixed at 10. In Hete L, data for each client $k$ are independently generated from $\mathcal{N}(\mu_k, 1)$, where $\mu_k \sim \mathcal{N}(0,1)$. In Hete D, data are generated from $\mathcal{N}(0,1)$, $\mathcal{U}(-1,1)$, and $\mathcal{C}(0,1)$ across different clients. ``hetero'' indicates client-specific truthful response rates $r_k$ range from $[0.25, 0.9].$}
\resizebox{0.8\textwidth}{!}{
\begin{tabular}{ccccccc}
\toprule
Quantile ($\tau$) & Rate ($r$) & C1 & C5 & Log &DP-SGD (C1) & DC \\
\midrule
\multicolumn{7}{c}{Hete L | $t_T = 10000$} \\
0.3 & 0.25 & 0.958(0.0184) & 0.981(0.0311) & 0.990(0.0452) & 0.942(0.0260)& 0.985(0.3066)\\
0.3 & hetero & 0.949(0.0096) & 0.982(0.0150) & 0.993(0.0205) & 0.947(0.0142)& 0.898(0.1302)\\
0.3 & 0.9 & 1.000(0.0029) & 1.000(0.0066) & 1.000(0.0100) & 0.981(0.0049)& 0.215(0.1273)\\
\hline
0.5 & 0.25 & 0.950(0.0165) & 0.984(0.0315) & 0.988(0.0465) & 0.953(0.0224)& 1.000(0.2822)\\
0.5 & hetero & 0.952(0.0085) & 0.991(0.0155) & 0.998(0.0221) & 0.955(0.0119)& 1.000(0.0525)\\
0.5 & 0.9 & 0.996(0.0025) & 0.999(0.0078) & 1.000(0.0120) & 0.984(0.0041)& 1.000(0.0186)\\
\hline
0.8 & 0.25 & 0.966(0.0237) & 0.995(0.0512) & 0.992(0.0791) & 0.957(0.0328)& 0.892(0.6152)\\
0.8 & hetero & 0.962(0.0122) & 0.995(0.0227) & 0.996(0.0347) & 0.943(0.0186)& 0.709(0.2684)\\
0.8 & 0.9 & 0.990(0.0042) & 1.000(0.0116) & 1.000(0.0185) & 0.968(0.0065)& 0.049(0.2098)\\
\hline
\multicolumn{7}{c}{Hete L | $t_T = 50000$} \\
0.3 & 0.25 & 0.937(0.0089) & 0.981(0.0111) & 0.990(0.0165) & 0.916(0.0135)&0.949(0.1328)\\
0.3 & hetero & 0.911(0.0056) & 0.981(0.0056) & 0.997(0.0080) & 0.885(0.0083)&0.093(0.1282)\\
0.3 & 0.9 & 0.977(0.0034) & 1.000(0.0019) & 1.000(0.0030) & 0.908(0.0041)&0.000(0.1290)\\
\hline
0.5 & 0.25 & 0.958(0.0069) & 0.988(0.0098) & 0.995(0.0147) & 0.949(0.0099)&1.000(0.0609)\\
0.5 & hetero & 0.964(0.0035) & 0.994(0.0048) & 0.996(0.0069) & 0.957(0.0052)&0.997(0.0145)\\
0.5 & 0.9 & 1.000(0.0010) & 1.000(0.0016) & 1.000(0.0026) & 0.993(0.0018)&0.979(0.0143)\\
\hline
0.8 & 0.25 & 0.956(0.0102) & 0.991(0.0144) & 0.998(0.0226) & 0.931(0.0160)&0.799(0.2829)\\
0.8 & hetero & 0.950(0.0055) & 0.992(0.0072) & 0.997(0.0112) & 0.923(0.0092)&0.014(0.2034)\\
0.8 & 0.9 & 1.000(0.0013) & 1.000(0.0053) & 0.999(0.0082) & 0.985(0.0026)&0.000(0.1929)\\
\midrule
\multicolumn{7}{c}{Hete D | $t_T = 10000$} \\
0.5 & 0.25 & 0.949(0.0132) & 0.985(0.0243) & 0.986(0.0354) & 0.953(0.0183) & 0.904(0.2496)\\
0.5 & hetero & 0.966(0.0069) & 0.990(0.0117) & 0.989(0.0172) & 0.955(0.0098) & 0.999(0.0488)\\
0.5 & 0.9 & 1.000(0.0023) & 1.000(0.0074) & 1.000(0.0117) & 0.991(0.0035) & 1.000(0.0163)\\
\hline
\multicolumn{7}{c}{Hete D | $t_T = 50000$} \\
0.5 & 0.25 & 0.958(0.0057) & 0.980(0.0082) & 0.993(0.0127) & 0.943(0.0081) & 0.981(0.0589)\\
0.5 & hetero & 0.966(0.0030) & 0.988(0.0046) & 0.998(0.0073) & 0.950(0.0041) & 1.000(0.0111)\\
0.5 & 0.9 & 0.999(0.0008) & 1.000(0.0029) & 1.000(0.0052) & 0.990(0.0014) & 1.000(0.0037)\\
\bottomrule
\end{tabular}
}
\label{tab:fixed n vary mu}
\end{table}

\begin{table}[htbp]
\centering
\caption{ECP (MAE) under varying quantile levels and response rates, with different $T$ and fixed $K = 10$ clients and data generated from $\mathcal{N}(0,1)$. In Case $\tau_{\operatorname{low}}$, each client uses a unique quantile level $\tau_k$ ranging uniformly from $[0.3, 0.5]$; in Case $\tau_{\operatorname{high}}$, $\tau_k$ is ranging from $[0.5, 0.8]$. ``hetero'' indicates client-specific truthful response rates $r_k$ range from $[0.25, 0.9].$}
\label{tab:fixed T vary tau}
\resizebox{0.8\textwidth}{!}{
\begin{tabular}{ccccc}
\toprule
Quantile ($\tau$) & Rate ($r$) & C1 & C5 & Log  \\
\midrule
\multicolumn{5}{c}{$T = 5000$} \\
0.5 & 0.25 & 0.954(0.0189) & 0.974(0.0129) & 0.986(0.0112) \\
0.5 & hetero & 0.959(0.0103) & 0.976(0.0065) & 0.995(0.0052) \\
0.5 & 0.9 & 0.999(0.0033) & 1.000(0.0040) & 1.000(0.0026) \\
\hline
$\tau_{\operatorname{low}}$ & 0.25 & 0.957(0.0200) & 0.974(0.0137) & 0.991(0.0116) \\
$\tau_{\operatorname{low}}$ & hetero & 0.957(0.0108) & 0.977(0.0067) & 0.993(0.0053) \\
$\tau_{\operatorname{low}}$ & 0.9 & 1.000(0.0033) & 1.000(0.0040) & 1.000(0.0029) \\
\hline
$\tau_{\operatorname{high}}$ & 0.25 & 0.956(0.0212) & 0.975(0.0128) & 0.993(0.0123) \\
$\tau_{\operatorname{high}}$ & hetero & 0.961(0.0112) & 0.984(0.0062) & 0.996(0.0056) \\
$\tau_{\operatorname{high}}$ & 0.9 & 0.998(0.0037) & 0.997(0.0028) & 1.000(0.0031) \\
\hline
\multicolumn{5}{c}{$T = 10000$} \\
0.5 & 0.25 & 0.949(0.0133) & 0.968(0.0078) & 0.987(0.0061) \\
0.5 & hetero & 0.963(0.0071) & 0.978(0.0037) & 0.991(0.0030) \\
0.5 & 0.9 & 0.995(0.0023) & 0.999(0.0020) & 0.999(0.0014) \\
\hline
$\tau_{\operatorname{low}}$ & 0.25 & 0.947(0.0136) & 0.972(0.0078) & 0.984(0.0064) \\
$\tau_{\operatorname{low}}$ & hetero & 0.962(0.0072) & 0.985(0.0038) & 0.983(0.0033) \\
$\tau_{\operatorname{low}}$ & 0.9 & 0.999(0.0020) & 1.000(0.0016) & 0.967(0.0018) \\
\hline
$\tau_{\operatorname{high}}$ & 0.25 & 0.939(0.0145) & 0.974(0.0086) & 0.985(0.0066) \\
$\tau_{\operatorname{high}}$ & hetero & 0.968(0.0076) & 0.988(0.0043) & 0.985(0.0032) \\
$\tau_{\operatorname{high}}$ & 0.9 & 0.996(0.0023) & 0.999(0.0031) & 0.996(0.0014) \\
\bottomrule
\end{tabular}
}
\end{table}

\begin{table}[htbp]
\centering
\caption{ {\small ECP (MAE) under heterogeneous distributions for different $T$. The number of clients $K$ is fixed at 10. In Hete L, data for each client $k$ are independently generated from $\mathcal{N}(\mu_k, 1)$, where $\mu_k \sim \mathcal{N}(0,1)$. In Hete D, data are generated from $\mathcal{N}(0,1)$, $\mathcal{U}(-1,1)$, and $\mathcal{C}(0,1)$ across different clients. ``hetero'' indicates client-specific truthful response rates $r_k$ range from $[0.25, 0.9].$}}
\label{tab:fixed T vary loc}
\resizebox{0.58\textwidth}{!}{
\begin{tabular}{ccccc}
\toprule
Quantile ($\tau$) & Rate ($r$) & C1 & C5 & Log  \\
\midrule
\multicolumn{5}{c}{Hete L | $T = 5000$} \\
0.3 & 0.25 & 0.942(0.0271) & 0.960(0.0168) & 0.975(0.0151) \\
0.3 & hetero & 0.962(0.0131) & 0.966(0.0086) & 0.987(0.0067) \\
0.3 & 0.9 & 0.998(0.0043) & 0.959(0.0063) & 1.000(0.0033) \\
\hline
0.5 & 0.25 & 0.954(0.0254) & 0.973(0.0154) & 0.990(0.0153) \\
0.5 & hetero & 0.963(0.0120) & 0.981(0.0072) & 0.991(0.0071) \\
0.5 & 0.9 & 0.992(0.0042) & 0.998(0.0032) & 1.000(0.0034) \\
\hline
0.8 & 0.25 & 0.954(0.0375) & 0.982(0.0242) & 0.998(0.0248) \\
0.8 & hetero & 0.968(0.0181) & 0.988(0.0109) & 0.998(0.0116) \\
0.8 & 0.9 & 0.985(0.0108) & 0.999(0.0070) & 0.982(0.0094) \\
\hline
\multicolumn{5}{c}{Hete L | $T = 10000$} \\
0.3 & 0.25 & 0.958(0.0184) & 0.966(0.0102) & 0.981(0.0083) \\
0.3 & hetero & 0.949(0.0096) & 0.965(0.0050) & 0.979(0.0040) \\
0.3 & 0.9 & 1.000(0.0029) & 0.979(0.0022) & 0.867(0.0036) \\
\hline
0.5 & 0.25 & 0.950(0.0165) & 0.974(0.0094) & 0.985(0.0085) \\
0.5 & hetero & 0.952(0.0085) & 0.976(0.0045) & 0.991(0.0039) \\
0.5 & 0.9 & 0.996(0.0025) & 0.985(0.0018) & 1.000(0.0016) \\
\hline
0.8 & 0.25 & 0.966(0.0237) & 0.983(0.0163) & 0.990(0.0149) \\
0.8 & hetero & 0.962(0.0122) & 0.988(0.0088) & 0.974(0.0090) \\
0.8 & 0.9 & 0.990(0.0042) & 0.997(0.0087) & 0.645(0.0095) \\
\hline
\multicolumn{5}{c}{Hete D | $T = 5000$} \\
0.5 & 0.25 & 0.954(0.0195) & 0.974(0.0129) & 0.987(0.0109) \\
0.5 & hetero & 0.965(0.0098) & 0.974(0.0075) & 0.993(0.0049) \\
0.5 & 0.9 & 1.000(0.0037) & 0.989(0.0060) & 1.000(0.0026) \\
\hline
\multicolumn{5}{c}{Hete D | $T = 10000$} \\
0.5 & 0.25 & 0.949(0.0132) & 0.968(0.0078) & 0.982(0.0064) \\
0.5 & hetero & 0.966(0.0069) & 0.973(0.0039) & 0.972(0.0034) \\
0.5 & 0.9 & 1.000(0.0023) & 0.999(0.0014) & 0.966(0.0023) \\
\bottomrule
\end{tabular}
}
\end{table}
\newpage
    
{\color{black}
\subsection{Real data}
In this subsection, we empirically evaluate the effectiveness of our proposed method using a representative real‐world dataset widely employed in privacy research: Government Salary Dataset \citep{plevcko2024fairadapt}. This dataset is sourced from the 2018 American Community Survey conducted by the U.S. Census Bureau and contains over 200,000 records, with annual salary (in USD) as the response variable. Since annual salary represents sensitive financial information \citep{pmlr-v139-gillenwater21a}, we treat it as requiring privacy protection. To incorporate the dataset's inherent geographic structure, we partition the sample according to the feature ``economic region." The three smallest regions are merged into a single ``Others" category, yielding seven regions in total, each regarded as one client. Because region-level sample sizes vary, we apply oversampling to balance the data, resulting in $t_T = 53,960$ observations per client. All other hyperparameters follow the settings in Section 4.1. For analysis, we apply a log transformation to the response variable and subsequently back-transform it.

We target quantile levels $\tau_k \equiv \tau \in \{0.3,0.5,0.8\}$ and consider response rate ranges from 0.6 to 0.9. For reference, we also compute the full-sample quantiles without LDP. The resulting estimators and confidence‐interval lengths are summarized in Table \ref{tab:real-data-results}. As shown, higher response rates $r$ and more communication rounds generally produce shorter confidence intervals, consistent with our simulation findings. In most cases, the empirical quantiles fall within our reported intervals, highlighting the practical utility of our method for real data.

\begin{table}[h!]
  \centering
  \caption{Estimation results (interval lengths) on the real dataset across varying quantile levels and response rates. “Empirical” denotes the full-sample quantile estimator without LDP. “hetero” indicates client-specific truthful response rates $r_k$ range from $0.6$ to $0.9$.}
  \label{tab:real-data-results}
  \begin{tabular}{ccccccc}
    \toprule
    Quantile ($\tau$) & Rate ($r$)    & C1            & C5            & Log           & Empirical \\
    \midrule
    0.3      & 0.6    & 33367 (1742) & 33184 (6697) & 33030 (12093) &      \\
    0.3      & hetero &  33418 (1424) & 33229 (5291) & 33140 (9788)  &  34000         \\
    0.3      & 0.9    &  33547 (1548) & 33403 (4443) & 33239 (7828)  &           \\
    \addlinespace
    0.5      & 0.6    & 48454 (2255) & 48212 (6315) & 47951 (11361)  &     \\
    0.5      & hetero &  48462 (1435) & 48290 (4973) & 48091 (9025)  &  50000          \\
    0.5      & 0.9    &  48610 (1454) & 48494 (3851) & 48311 (6863)  &           \\
    \addlinespace
    0.8      & 0.6    & 78586 (2066) & 78168 (6646) & 77995 (13144)  &    \\
    0.8      & hetero & 78390 (1291) & 78054 (5862) & 77722 (11101)  &  80000           \\
    0.8      & 0.9    &  78657 (1138) & 78300 (4677) & 78084 (8928)  &           \\
    \bottomrule
  \end{tabular}
\end{table}

}
\section{Concluding remark}
\par We propose a federated‐learning algorithm for quantile inference under LDP that flexibly accommodates client‐level heterogeneity in quantile targets, privacy budgets, and data distributions. In addition, one innovation that should be emphasized is that our developed theoretical results of local SGD quantile estimator. We first design an LDP mechanism that can transform the LDP federated quantile estimation into the non-DP case, and then derive the asymptotic normality and functional central limit theorem of the proposed estimator under non-DP cases. It is first weak‐convergence result for local SGD without the usual average‐smoothness assumption in existing literature. Building on these non‐private asymptotic results, we develop a self‐normalized inference procedure that constructs valid confidence intervals under LDP without requiring direct estimation of the asymptotic variance.

\par Despite these advances, our method has several limitations. First, it relies on additional regularity assumptions to handle arbitrary client‐level data heterogeneity. Second, as noted in \citep{shao2015self}, self‐normalization yields heavier‐tailed limit distributions than the Gaussian, which can produce conservative confidence intervals or reduced power in hypothesis testing. Finally, our framework depends on a central server for aggregation and synchronization, which may not be available in fully decentralized environments. Addressing these challenges and extending the algorithm to decentralized settings remain important directions for future research.

\bibliographystyle{apalike}
		\bibliography{ref}

\end{document}